# Predicting the real-valued distances between residue pairs for proteins

*Wenze Ding[1,2] and Haipeng Gong[1,2,*]*

[1]MOE Key Laboratory of Bioinformatics, School of Life Sciences, Tsinghua University, Beijing 100084, China.

[2]Beijing Advanced Innovation Center for Structural Biology, Tsinghua University, Beijing 100084, China.

*To whom correspondence should be addressed:

Email: hgong@tsinghua.edu.cn (H. G.)

**Abstract:** Predicting protein structure from the amino acid sequence has been a challenge with theoretical and practical significance in biophysics. Despite the recent progresses elicited by improved residue-residue contact prediction, contact-based structure prediction has gradually reached the performance ceiling. New methods have been proposed to predict the residue-residue distance, but unanimously by simplifying the real-valued distance prediction into a multiclass classification problem. Here we show a regression-based distance prediction method, which adopts the generative adversarial network to capture the delicate geometric relationship between residue pairs and thus could predict the continuous, real-valued residue-residue distance satisfactorily. The predicted residue distance map allows rapid structure modeling by the CNS suite, and the constructed models approach at least the same level of quality as the other state-of-the-art protein structure prediction methods when tested on available CASP13 targets. Moreover, this method can be used directly for the structure prediction of membrane proteins without transfer learning.

# 1. Introduction

Proteins participate in nearly all kinds of physiological activities and their three-dimensional structures are essential for the functions. Since the finding that the protein structures are prescribed by their amino acid sequences, exploration of the relationship among protein sequence, structure and function has been one of the core problems of molecular biophysics. Unlike experimental structure determination methods that are costly and technically prohibitive, predicting the protein structure via computational approaches could be applied in a high-throughput manner and thus is widely needed in practical applications ranging from protein design to pharmaceutical development[1].

In traditional *de novo* protein structure prediction, the native structure is located by exhaustively searching the protein conformational space, using molecular dynamics simulations that employ empirical force fields or fragment-assembly-based/threading-assembly-based Monte Carlo simulations that use experimentally determined structures as templates and force fragments of the target protein to adopt their conformations[2]. Despite the successes, traditional methods become less powerful for hard protein targets that have complex topologies but limited homology to known structures, e.g., the free-modeling (FM) targets in the Critical Assessment of protein Structure Prediction (CASP) competitions. In addition, these methods are computationally expensive in general, because the protein conformational space frequently has intimidatingly high dimensionality.

Breakthrough in the accuracy of protein structure prediction was observed in CASP11 and CASP12[3-5] (hosted in 2014 and 2016, respectively), which was mainly driven by the use of co-evolution information and deep learning algorithms. For a target protein, evolutionary couplings between residues could be detected and extracted from the multiple sequence alignment (MSA) to predict the binary contact matrix, assuming that spatial neighborhood of residues (so-called residue contact, strictly defined as the distance of $C_\beta$ atoms ≤ 8 Å according to the CASP convention) would elicit correlated mutations over long evolutionary time. The contact matrix contains geometric constraining information that could be used by protein folding programs such as CONFOLD[6] to restore the atomic coordinates. On the other hand, traditional methods also benefit substantially from the integration of contact prediction in structure selection and energy evaluation[7]. Extraction of contact information from the MSA is a typical pattern recognition problem that is particularly suitable to be handled by deep learning techniques like convolution neural network (CNN)[8], because of the power of such techniques to identify the correlation between contacting residue pairs located far away in the contact map. Among many deep-learning-based approaches, RaptorX-Contact that adopts deep residual network (ResNet) outperforms others and makes huge influences in this field[9].

Nevertheless, contact prediction is just a compromise when accurate distance prediction is not available. Distance prediction has many intrinsic advantages over contact prediction for protein folding. First of all, contact prediction is a binary classification problem with unbalanced positive and negative samples (e.g., roughly 1 contact : 50 non-contacts for long-range residue pairs with sequence separation ≥ 24)[10], which frequently requires undersampling of negative ones during model training and thus may lead to the inconsistence between the prediction score and the real contacting probability for a residue pair. Therefore, contact-assisted protein folding methods usually only adopt a small parts of predicted contacts with top scores for structure modeling, which is susceptible to the noises raised by very few wrongly predicted contacts. Direct prediction of distance map (i.e. the 2D matrix listing real-valued distances between all residue pairs) would avoid this problem, because all predicted values within a suitable interval (e.g., 4-16 Å) can be utilized and thus the disturbance introduced by the prediction errors of individual residue pairs may mitigate according to the law of large number. More importantly, distance matrices contain much more detailed information of protein structure than contact matrices, which could reduce conformational sampling more effectively and thus fold the protein more accurately and rapidly. Consequently, despite the great progresses introduced by contact prediction, first-ranked groups in the field of protein structure prediction like AlphaFold and RaptorX-Contact have switched their attention to distance prediction in CASP13[9, 11] (hosted in 2018).

Ideally, during the switch from contact prediction to distance prediction, the nature of the explored task should transit from a classification problem to a regression problem, because residue contacts are actually human-defined zero-one labels while distances are real-valued physical metrics. However, both AlphaFold and RaptorX-Contact simply chose to modify binary classification to multiclass classification. Instead of real-valued distances, they used a discrete representation with several fixed-width bins[9, 11]. The rationale for their choices is mainly 3-fold. Firstly and foremost, traditional regression loss functions used in deep neural network (DNN) like mean absolute error (MAE, also called L1 loss) and mean square error (MSE, also called L2 loss) measure the globally averaged deviation of the prediction from the ground truth. After loss minimization by DNN, the predicted distances may be pretty good on average but still far from satisfaction as individuals, which is of limited usefulness for protein folding. Generally, it is hard and needs many manual efforts to design effective losses for separate, special-purpose machinery. Secondly, modern DNN training procedures always add batch normalization (BN) layers to solve the problem of gradient vanishing or explosion, which normalize the forward-passing data into a standard normal distribution. Thus, without ingenious, human-designed mapping functions, common activation functions used in DNN are powerless of outputting positive real numbers like distances. Thirdly, with their powerful, well-tested contact prediction networks in hand, these groups can conveniently use transfer learning techniques to get satisfactory distance prediction results.

In this work, we solved all the obstructions for distance prediction via exquisitely designed generative adversarial networks (GANs) and directly predicted the real-valued residue-residue distances with satisfactory accuracy for the first time. GAN is a computer vision technique, containing a generator to produce outputs and a discriminator to classify outputs of the generator from the real ones (ground truths). With joint training, the generator would not only fit the pixel distribution of real image globally, but also highlight "important" pixel areas with sharp, precise values to fool the discriminator simultaneously[12, 13, 14]. Other contributions of our work include (1) introducing new, effective data augmentation methods to produce more robust models, especially the augmentation of distance labels by molecular dynamics simulations, which considers the structural dynamics of proteins that are typically ignored in structural bioinformatics, (2) designing reversible mapping functions between positive real numbers and the interval of [-1, 1] to enable the direct training of DNN for continuous residue-residue distance regression, and (3) analyzing the effects of several technical choices and then summarizing some empirical laws for the deep learning solution of residue-residue distance prediction for proteins. When pipelined with the same protein folding program CNS suite[15], structure models generated using our distance constraints are significantly better than those produced with the contact constraints from the state-of-the-art contact predictors like RaptorX-Contact[9] and TripletRes[16]. Moreover, when tested on available CASP13 targets, our structure models approach at least the same level of quality as the top protein structure prediction groups, including AlphaFold (A7D)[11], Zhang[17] and MULTICOM[18]. Although trained mainly by protoplasmic soluble proteins, the generalizability of our predictor renders its application for the structure prediction of membrane proteins without the requirement of any transfer learning processes.

## 2. Results

### 2.1. A preliminary GAN model for protein distance prediction

We first developed mapping functions to allow the back and forth transformation between real-valued distances or features and numbers in [-1, 1] (see Experimental Section for details), through which the ground-truth distance maps could be converted to the interval of [-1, 1] to simplify the training of DNN models with BN layers and the prediction results could also be restored to the domain of real distances instantly. Particularly, the mapping functions were designed to have large gradients for distances between 4 and 16 Å, the range possessing rich information for the protein structure modeling. We then adopted the conditional GAN (cGAN) for protein distance prediction. Similar to but distinct from primitive GANs, cGAN learns a generative model (generator, referred as G) that can generate the corresponding output of expected size in the condition of an input[12]. Here, a 40-layer ResNet, one of the most successful network architectures in this field[9], was chosen as the generator of our cGAN and was also taken as the control to evaluate the performance gain of GANs over pure generative models (**Figure 1a**). The discriminator of our cGAN, referred as D, is trained to detect the outputs of G as "fake" from "real" under the condition of input features fed to G, whereas G tries to learn from the decision of D and produces indistinguishable outputs to "fool" D through the adversarial training procedure. More specifically, the loss function of D can be defined as a standard cross-entropy function for a binary classifier with a sigmoid output:

$$LossD = -LossGAN(G,D) = -(E_{x,y}[logD(x,y)] + E_{x,z}[log(1 - D(x, G(x,z)))]), \quad (1)$$

where $x$, $y$ and $z$ represent input features, real distance maps and input noises, respectively, and $E$ denotes expectation. Notably, unlike the common GANs that apply noises to ensure the randomness of outputs, we did not increase noises in G, because G in our experiments is robust such that it learns to ignore any kind of tiny disturbance. D tries to minimize **Equation 1** against the adversarial G, which in return tries to maximize it (i.e. minimize its negative number, *LossGAN*). Besides fooling D, G should also constrain its outputs near the ground truths. Hence, it would be beneficial to combine a more traditional regression loss (*RegLoss*), and the final G loss is defined as

$$LossG = LossGAN(G,D) + \lambda \times RegLoss, \quad (2)$$

where $\lambda$ is a weight parameter to adjust the relative importance of two parts. In this section, we chose L1 loss as the regression loss and 258 for its weight. For the consistence between cGAN and the control, L1 loss of the control ResNet was also multiplied by the same $\lambda$:

$$LossControl = \lambda \times RegLoss. \quad (3)$$

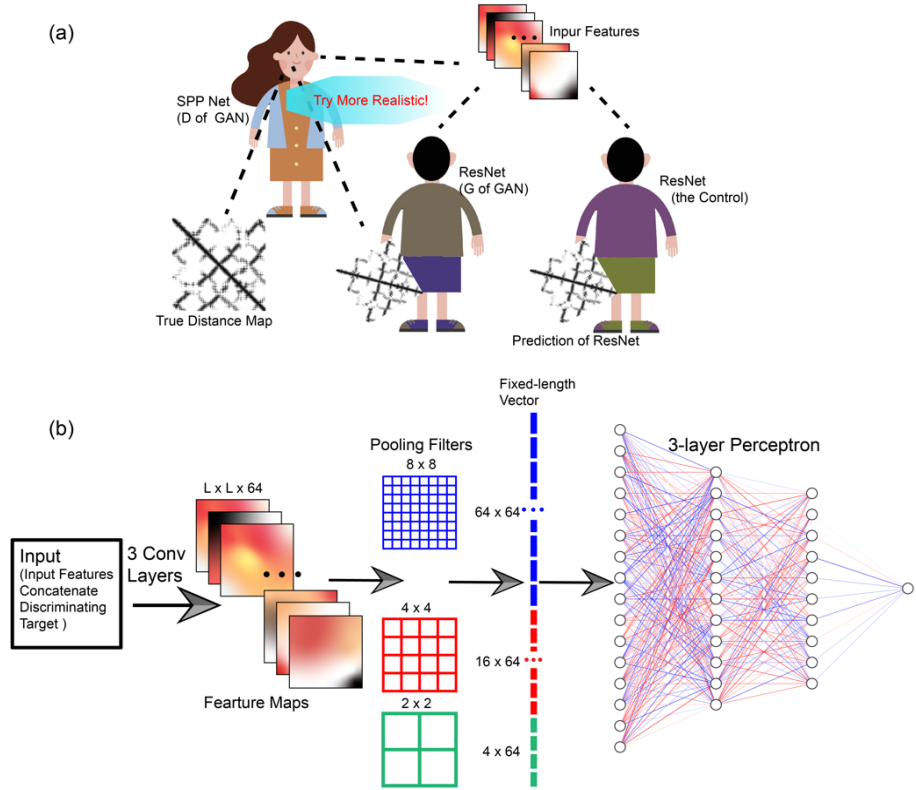

**Figure 1.** Schematic illustrations of the preliminary experiment. **(a)** D and G of our GAN model contest as adversaries to improve the quality of predicted distance map, in comparison with the traditional ResNet as control. The dished lines represent information flows. **(b)** The architecture of the D model.

We set each individual protein as a mini-batch during training. For G of our cGAN and the control ResNet, 64 3×3 2D convolution filters with stride 1 and zero-padding "same" were adopted for each layer, followed by the leaky rectified linear unit (leaky-ReLU) and BN. For D, we concatenated the input features of G and the discriminating targets (i.e. outputs of G or ground truths) as its input. To solve the problem of variable sizes of individual proteins, we adopted spatial pyramid pooling (SPP)[19] following the 3 convolutional layers (each with 64 3×3 2D filters) in D, where the max-pooling results of three different separations (8×8, 4×4 and 2×2 patches) of the feature map were concatenated as a fixed-length vector and were fed into a 3-layer perceptron to output the probability of the given distance map to be true (**Figure 1b**). The training procedures of cGAN and the control ResNet were completely the same, using the Adam optimizer for 100 epochs with the learning rate set as 1e-4, 1e-5 and 1e-6 for the first 20, the middle 30 and the last 50 epochs, respectively. We randomly chose 5642 proteins from the dataset of 6862 chains as the training set, and left the rest 1220 proteins as the validation set. To speed up training, the maximal length of proteins was limited to 400 residues.

**Table S1** summaries the prediction errors by the cGAN and the control ResNet in the validation set. For residue pairs with the predicted distances falling between 4 and 16 Å (the range having rich information for protein structure modeling), ResNet seems to reach lower prediction error than cGAN on average (1.832 Å vs. 1.938 Å). However, are the "seemingly better" results by ResNet really benefiting the protein structure modeling? To address this question, we collected all predicted distances within 4 and 16 Å to construct the distance constraining matrix and invoked the CNS suite[15] (using a similar protocol to CONFOLD[6]) to fold the proteins in the validation set. To ensure that CNS suite indeed uses the predicted residue distances for structure modeling, we chose a narrow distance range of ± 0.4 Å around the predicted value. Quality of the top 1 models was evaluated by TM-score. As shown in **Figure 2a**, cGAN defeats the control ResNet for most targets in validation set. The models folded by cGAN predictions reach an average TM-score of 0.722, with 92.7% of the targets folded in the correct topology (i.e. TM-score > 0.5). In contrast, the average TM-score of ResNet-based folding is 0.544, with only 63.9% of the targets folded correctly. Hence, despite the slight weakening of the overall distance prediction accuracy, introduction of the GAN loss that comprehensively considers the adversarial generator and discriminator (see **Equations 1&2**) indeed improves the structure modeling based on the predicted distances.

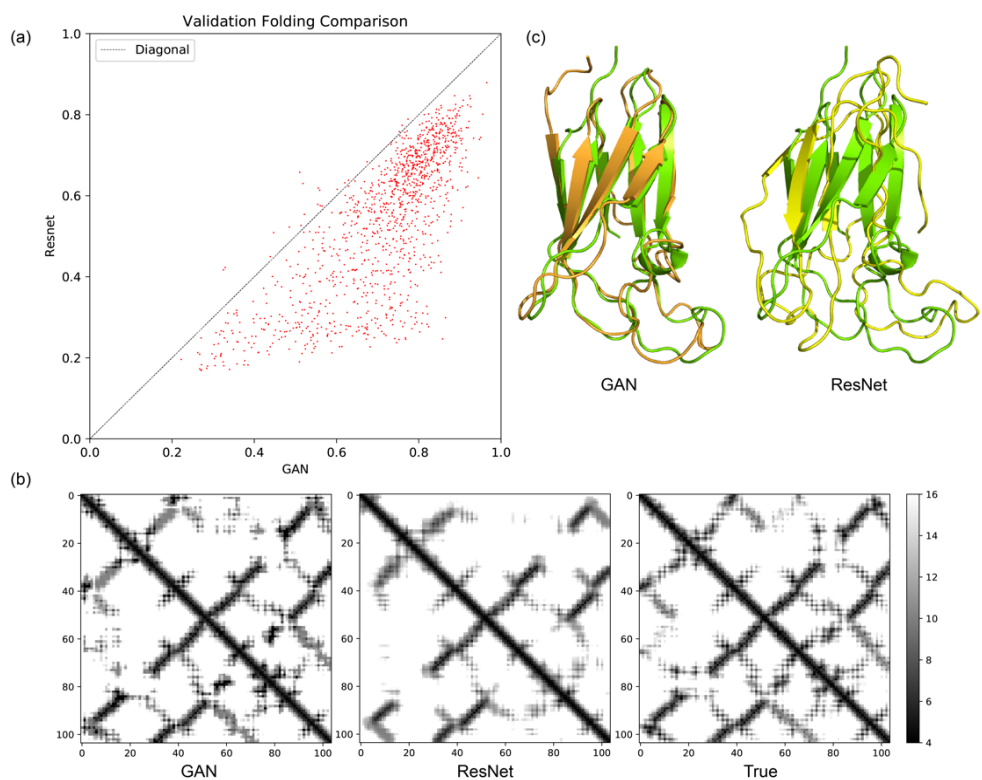

**Figure 2.** Comparison between our cGAN system and ResNet (the control) for protein structure prediction. **(a)** Comparison of the TM-scores of the structure models produced by CNS-based folding using the distance predictions of cGAN and ResNet as restraints in the validation set. **(b)** Predicted distance maps by cGAN (left) and ResNet (middle) are compared with the true

distance map (right) for a case target (PDB ID: 2II8). The bar on the right indicates the grayscale for the predicted distance (Å). **(c)** Structure alignment of the best folded models using cGAN (orange, left) and ResNet (yellow, right) predictions against the crystal structure (green).

**Figure 2b** shows the distance maps predicted by cGAN and ResNet as well as the ground truth for an example target (PDB ID: 2II8). Clearly, the prediction by ResNet is blurry overall, although locations and average values of the main stripes are roughly correct. In contrast, despite many tiny mistakes, the prediction by cGAN contains much more details with sharp edges. The sharp contrasts between pixel signals captured by cGAN prediction describe the subtle correlations between individually predicted distances, which imply the delicate geometric relationship between residue pairs. Consequently, the structure model generated by cGAN prediction agrees with the native structure significantly better than that by ResNet prediction (**Figure 2c**).

The fitting power of ResNet is guaranteed by the multiple stacking of residual blocks even when the size of convolution filter is small, as the receptive field would be amplified in a cascaded way and thus the interdependency of two arbitrary residue pairs could be captured. Thus, the question is focused on what we want our neural network to fit, and since the network learns to minimize a loss function that evaluates how close the outputs of the network and our desires are, the question finally becomes how we define the loss of our neural network. It is well known that traditional regression loss like L1 or L2 losses could capture the low-frequencies, i.e. average information, accurately from inputs. They measure the global quality of outputs and thus drive the networks to produce predicting values around the local average, which as a result may blur their outputs. However, these accurate low-frequencies are far from the demand of practical usage in protein folding, and what we really want is a realistic residue distance map with sharp contrasts between pixels. Designing effective losses specifically for the extraction of these high-frequencies, i.e. texture information, is difficult because the high frequencies somehow represent the general properties of polypeptides or the protein folding mechanism like the interacting pattern between secondary structure elements and the local folding propensity of loop regions. Avoiding directly defining such kind of texture losses, our GAN solved this problem through achieving a high-level goal of "producing reality-indistinguishable predictions" and used a neural network D to learn this loss. At the same time, our GAN trained its generative model G to minimize the learned loss, which successfully suppressed the unrealistic blurs and reproduced high-frequencies.

## 2.2. Introducing patch classifiers to the architecture of D

As a data augmentation method, cropping has been proved as useful in practice by many research groups in this field. For example, AlphaFold randomly chose 64×64 patches from the protein feature map when training their 660-layer ResNet, which brought about many benefits, such as solving inconsistency problem of protein length variation, helping distributed training, avoiding overfitting problem and facilitating ensemble average for inference[11]. However, direct imitation of such cropping in our case failed in the GAN training.

Markovian discriminator was proposed recently to model high-frequencies in GAN[14]. Instead of determining whether the entire output is "real" or not, such kind of discriminators pay attention to subtle structure differences in output patches of fixed size. Inspired by this idea, we implemented our patch classifier as an alternative cropping method in D through a fully convolutional network (FCN). Each layer of this FCN adopts the 4×4 convolution kernel with the leaky-ReLU set as activation and zeros padded around inputs when necessary. The kernel stride of precedent layers was set as 2 to enlarge receptive field rapidly while the stride of the last two layers was set to 1 to better integrate information captured in each neuron. The channel number of the first convolution layer was set as 128, and the following ones were doubled at each turn except the last layer, where the channel number was set to 1. Sigmoid function was used as activation for the last convolution layer to output the probability of the corresponding patch to be true. The patch size could be modified with the depth variation of this FCN. For example, as shown in **Figure 3**, if we want the classifier to focus on 34×34 patches, the FCN should have 4 layers totally, with strides of (2, 2, 1, 1) and channel numbers of (128, 256, 512, 1). Such dense sampling of patches ensures coverage of the whole distance map without omission.

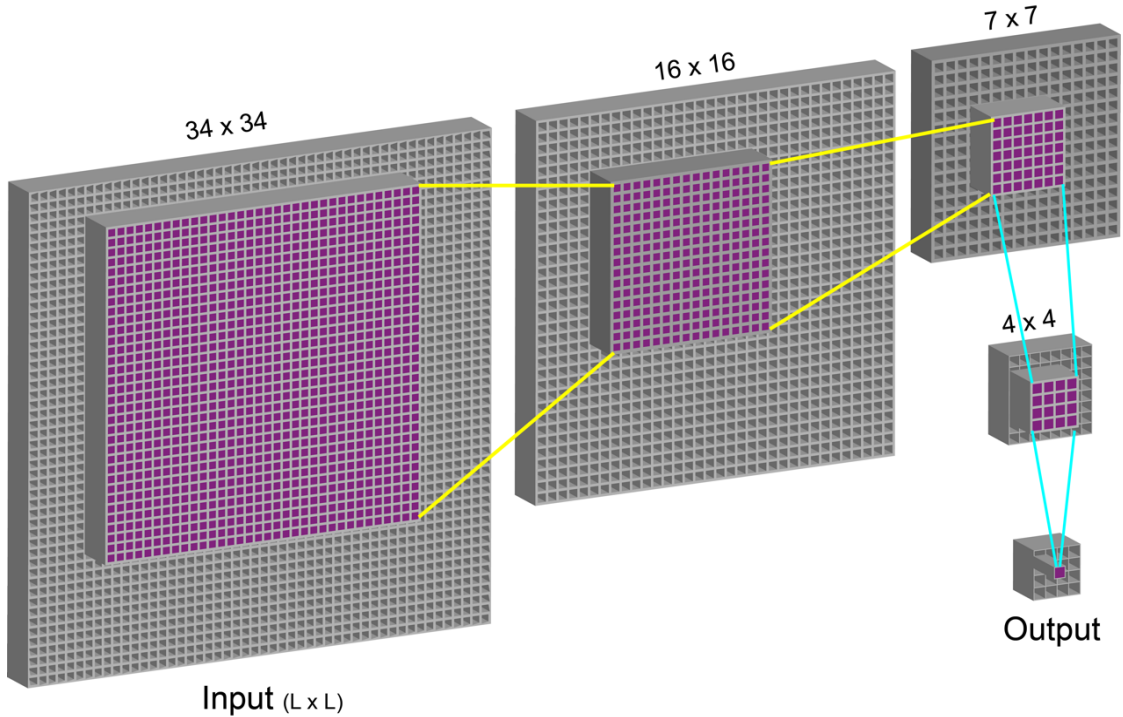

**Figure 3.** Illustration of the patch classifiers (34×34) in D. The size of patch (purple and highlighted as convex) evolves along sequential convolutional layers, with the size of feature map (gray) changing accordingly. Yellow and blue lines represent convolutional strides of 2

and 1, respectively. Numbers on top of the feature maps list the patch size in each layer.

We observed that the patch classifiers that have fewer parameters and faster speed indeed produced better results than the single classifier that makes judgement on the entire input of distance map (**Figure S1**). Because residue pairs separated by a patch diameter or longer intervals are frequently independent considering the statistical length of protein secondary structure elements (i.e. helices and sheets), patch classifiers could model the residue distance map as a Markov random field. Thus, the loss learned by patch classifiers should be useful for the extraction of the special texture pattern of distance map.

The characteristic of distance map hints us that we should pay more attention to those patches having stripes of strong signals (i.e. predicted distance between 4-16 Å) since only such predictions are meaningful and contributive to protein folding in our case. However, the distance map is usually dominated by blank background regions (i.e. predicted distance > 16 Å), which albeit lacking useful information are highly likely to be judged by D as "real" because they seem identical to the corresponding blank patches on the ground-truth map (**Figure S2**). To reduce such confusion of D, we modified its cross-entropy loss from **Equation 1** to

$$LossD = -\left(E_{x,y}[logD(x,y)] + E_x\left[log\left(CLIP(0.9 - D(x, G(x)), 0, 0.9)\right)\right]\right), \quad (4)$$

where *CLIP(fun(a),0,0.9)* is a clip function that only retains values of *fun(a)* within the 0-0.9 interval. Through this modification, patches seeming realistic (e.g., those blank background regions without stripes) at the first beginning when G has not learned useful information would be filtered out for the decision of D.

## 2.3. Optimizing the generative model G

As G is the major undertaker for information integration and extraction from inputs and the actually used part during inference, its architecture is vital to performance of the overall network. In this section, we adjusted all components of G one after another and summarized some empirical laws of the technical choices for distance prediction. All effects of adjustments were analyzed under rigorous 5-fold cross validation on the protein dataset of 6862 chains.

During training, we frequently observed the premature convergence of the GAN loss of G (*LossGAN(G,D)* in **Equation 2**). This is because D is likely to reject all the distance maps produced by G with high confidence after a few epochs when G cannot really learn something, considering that the regression task of G is much harder than the classification task of D. Albeit correct, such rejections quickly reduce the loss of D (**Equation 1**), which prevents the further learning by G. To solve it, we modified the GAN loss of G to favor the cases when D accepts distance maps predicted by G as "real"

ones:

$$LossG = E_x[-logD(x, G(x))] + \lambda \cdot RegLoss. \quad (5)$$

We tried a number of common network architectures for G, including different ResNet variants, DenseNet, and U-Net (**Table S2**). Among them, U-Net was hard to implement for inputs of variable sizes by available convolutional depression and recovery techniques, and we had to pad zeros around the inputs to ensure the length uniformity, which impaired the performance, because the padded zone might be much larger than original size for many proteins. Taking computational consumption (i.e. amount of parameters and FLOPs) into account, the performance of DenseNet is not satisfactory. For ResNet, the 3-layer-per-block variant outperforms the 2-layer-per-block one. Bottleneck structure seems not beneficial to distance prediction because the 1×1 kernel is incapable of enlarging the receptive field. We finally picked up the ResNet architecture under the guidance of EfficientNet[20] and restricted the protein length to 300 to avoid memory overflow (one 2080Ti card, memory 10989MiB).

As for the convolution kernel of G, we tried different kernel sizes, kernels with dilation and separable kernels (**Table S3**). Larger kernels have better performance by considering more complex interdependencies. As a frequently used technique in contact prediction and distance prediction with multi-classifiers, dilated kernels underperform normal ones. The reason of this phenomenon is that unlike classification problem that only needs to learn categorical information, real-valued distance regression requires large receptive field without any omission, especially for the extraction of texture information. We finally chose the 7×7 kernel. Since the training of models with 7×7 kernels was relative slow, we adopted parameter sharing technique proposed in the work of ShaResNet[21] to reduce the amount of parameters and accelerate training. Unfortunately, this procedure led to training failure.

In the evaluation of activation functions, the Swish function

$$Swish(x) = x \cdot sigmoid(\beta x), \quad (6)$$

which has a learnable parameter $\beta$, achieves the best performance (**Table S4**). Generally, activations with tunable parameters outperform those without, because they can mimic real biological neural networks, in which every individual neuron has its own property and activation threshold. It is noteworthy that random-ReLU (R-ReLU) impairs the performance, which is inconsistent with our previous experience on protein contact prediction. Among the regression losses tested, the MAE loss (or L1 loss) is the simplest but the most effective one (**Table S5**). This is because instead of biasing predictions of larger ground truths that usually have larger errors, the L1 loss balances various kinds of predictions well and thus performs more robustly.

Among features from various sources, 2D features are the most valuable. However, unlike AlphaFold that uses enormous 2D features directly from the Potts model to ensure information coverage[11], 2D features in our model only occupy 3.07% of the inputs (4 out of 130, see Experimental Section). Although these features are extracted from the MSA and are thus informative, their contributions may be submerged by the large amount of redundant information produced by the broadcasting of 1D features. Besides, the unbalanced value distribution of the protein distance map (e.g., the sparse distribution of intense "stripes" on the vast blank background area) further complicates the training. Inspired by these, attention aiming to reweight the channels (i.e. different features) and pixels (i.e. different regions) of the input features in consideration of individual targets is necessary in our system. We implemented an attention module with global average and max pooling (see Experimental Section) and the results supported its effectiveness in improving the model performance (**Table S6**). In the validation set, the channel-wised attention sufficiently suppresses the weights of redundant information from the broadcasting of 1D features (**Figure S3**). Meanwhile, the pixel-wised attention effectively adjusts the weights of individual pixels to facilitate information extraction (**Figure S4**).

## 2.4. Data augmentation with biological significance

What kind of structures should the predictors in the bioinformatics field predict? It is an open question since the structures from the Protein Data Bank (PDB) that are used as ground truths for model training are static structures determined in non-physiological conditions. Different crystallization situations, different structure analysis technologies (NMR, X-Ray, cryo-EM, etc.) and even different structure computation methods may lead to structure variation. More importantly, these static structures are unable to reflect the dynamic behaviors of real proteins in aqueous environments. Molecular dynamics (MD) simulations could solve this problem, because physiological environments are constructed and empirical force fields are adopted to observe the protein dynamics starting from the PDB structure in these simulations. To guarantee the generalizability and robustness of our predictor as well as to consider the protein dynamics, we augmented data via MD simulations (see Experimental Section). For each protein in our training set, 500 structures were collected with even separation from the 5-nanosecond trajectory of equilibrium simulation. The conformational change reaches the RMSD level of 6 Å on average at the end of the simulation (**Figure S5**), which ensures that structural dynamics are sufficiently considered by our data augmentation. To validate the contribution of this data augmentation on practical protein structure prediction, we trained two models of the same architecture (L1 loss weight = 100, patch size of D = 70 and without clipping) using structures from PDB (referred as Model 1) and structures produced by MD simulations (referred as Model 2), respectively, and utilized their results to fold proteins in the CASP13 set and non-redundant membrane protein set by the CNS suite. Enhancement in the quality of folded structures by Model 2 supports that our data augmentation indeed captures something with biological significance, which improves the generalizability of distance prediction and benefits the distance-based structure modeling (**Table S7**).

As mentioned above, the weight of L1 loss in the loss of G (λ in **Equation 5**), patch size of D and whether to clip the loss of D (**Equation 4**) more or less affect the distance information extraction. To enhance the variation of our prediction results, we trained 14 models at different combinations of these options (**Table S8**). For each protein, we randomly chose 4 structures as ground truths from the 501 available structures (one PDB structure plus 500 simulation-produced structures) during the training of each individual model, which augmented training samples by 4 times (**Figure 4**). To stabilize the training of our GAN system, we used exponential moving average for loss when updating network parameters. Independent tests on non-redundant membrane proteins as well as CASP12 and CASP13 targets show the diversity and mutual complementarity of these models (**Tables S9-S12**). Because we did not apply any symmetry restriction on the outcomes, the distance map predicted by each model could be treated as 3 different ones: the upper triangle, the lower triangle and their average. The value distribution at each pixel of 3×14=42 predictions in total fits the lognormal distribution well. Moreover, previous work showed the benefits of lognormal distribution for distance-based protein structure determination[22]. Thus, we computed the mean and standard deviation for the distance of each residue pair from the fitted lognormal distribution, and the mean ± standard deviation could be used as boundaries of the allowed distance range to fold the protein by the CNS suite.

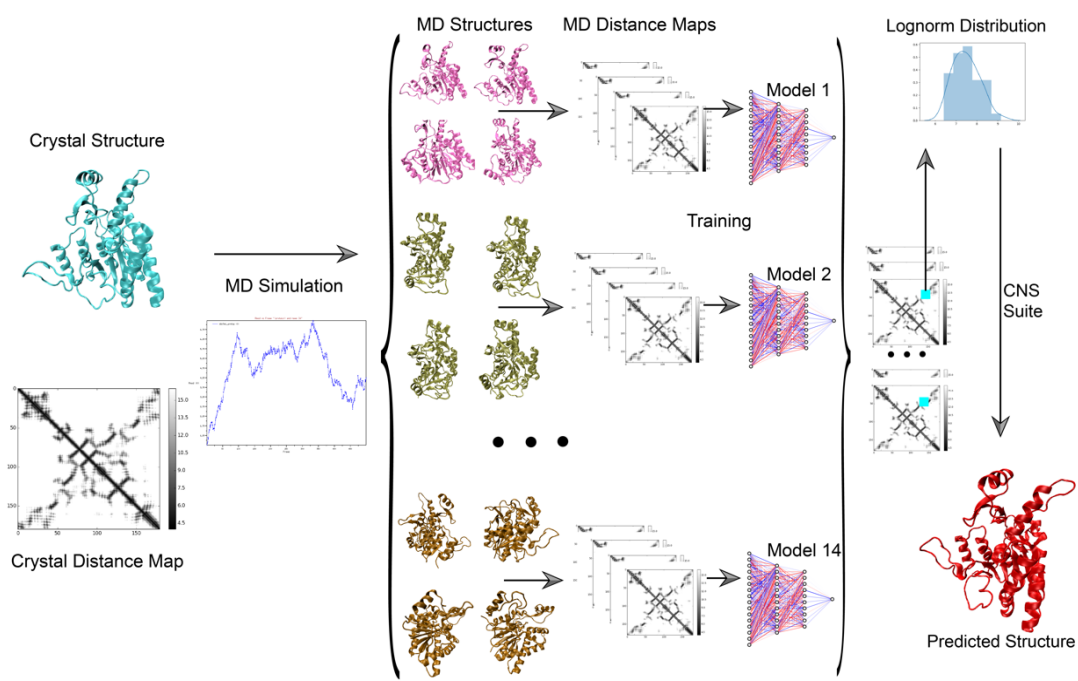

**Figure 4.** Flowchart of our training procedure, final inference and CNS-based folding. The blue dots marked on predicted distance maps represent a variety of prediction results at same pixel.

## 2.5. Evaluation

To validate the superiority of distance prediction to contact prediction, we compared our distance predictor with our contact predictor DeepConPred2[10] in the CASP12 set and with the top CASP13 contact predictors including RaptorX-Contact and TripletRes in the CASP13 set, by uniformly using the CNS suite to fold the proteins. The allowed distance range of each residue pair was set as the mean ± standard deviation derived from the lognormal distribution in our system as mentioned previously, but was set to [3.5, 8] for all contact predictors following the CONFOLD protocol. Clearly, our GAN system significantly outperforms DeepConPred2 that has similar input features (**Table S13**). Moreover, structure models constructed from our distance prediction have remarkably better quality than those generated based on the results of the state-of-the-art contact predictors (**Table 1** and **Figure S6**). The lead of our method by a large margin in TM-score supports the important contribution of real-valued distance prediction in protein structure prediction.

**Table 1.** Comparison of the folding capability by our distance predictor against contact predictors in the CASP13 set.

|  | Average TM-score | Correctly Folded |
| --- | --- | --- |
| TripletRes | 0.568 | 27 |
| RaptorX-Contact | 0.527 | 22 |
| Our GAN system | 0.719 | 31 |

"Correctly Folded" means the number of targets in the correct topology (TM-score > 0.5).

We also compared the structures folded using our distance prediction against the top 3 protein structure prediction groups in CASP13 (A7D, Zhang and MULTICOM) on 39 available CASP13 targets (T0951-D1, T0967-D1 and T0971-D1 were excluded because of missing records for these groups on the CASP13 homepage). As shown in **Table S14**, our method reaches an average TM-score of 0.70 for all targets, and the average TM-scores for FM targets and template-based-modeling (TBM) targets are 0.65 and 0.75, respectively. All of the above numbers are very close to the results of the top CASP13 groups. Particularly, our method achieves the highest average TM-score for FM targets that lack available structure templates and are thus more difficult for structure prediction.

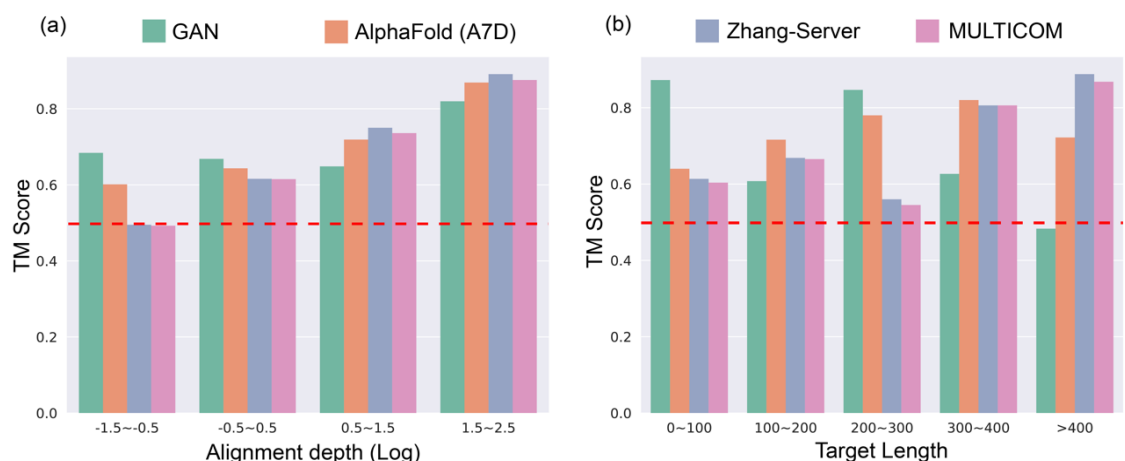

**Figure 5.** Comparison of TM-scores for targets by our method vs. the top CASP13 groups. **(a)** Targets are classified by logarithmic values of the alignment depth (*N/L*). **(b)** Targets are classified by their length. Red dashed line represents TM-score threshold of 0.5 for correctly folded targets.

**Figure 5** shows the detailed comparison of the TM-scores between our method and the top CASP13 groups, on targets of various alignment depths (i.e. *N/L*, where *N* is the effective number of homologous sequences and *L* is the protein length) and protein sizes. Clearly, our method outperforms the others for the hard targets that have low alignment depths in the MSA but become slightly less powerful for those easier targets, which is impressive particularly when considering that the others used structure-sourced information more or less and we used sequence information only. On the other hand, in comparison to the others, our method exhibits good performance for proteins with sizes of < 300 residues but becomes less powerful for larger proteins. This problem is mainly because those larger proteins were excluded from the training of our GAN system due to the limitation by GPU memory, but could be solved by training using more advance hardware (see Discussion). Nevertheless, the prediction results by our method exhibit a pattern considerably distinct from those of the other state-of-the-art methods, which implies its capability of providing complementary information in practical protein structure prediction. Moreover, it is noticeable that our GAN system pipelined with CNS suite could be deployed on personal computers with GPU cards, while the others required heavy computational resources.

Since all features used in our GAN system are sequence information extracted from the MSA, it is reasonable to assume that the prediction quality has correlation with the alignment depth, just as what happened in contact prediction and distance prediction using multi-classifiers. To check this point, we did correlation analysis on 42 CASP13 targets. The final prediction quality shows no significant correlation with the alignment depth of the target, with Pearson correlation coefficient (PCC) = 0.27 and p-value = 0.081 (**Figure 6**). More interestingly, for FM targets, the prediction quality becomes negatively correlated with the alignment depth instead, seemingly indicating that harder targets could be better predicted.

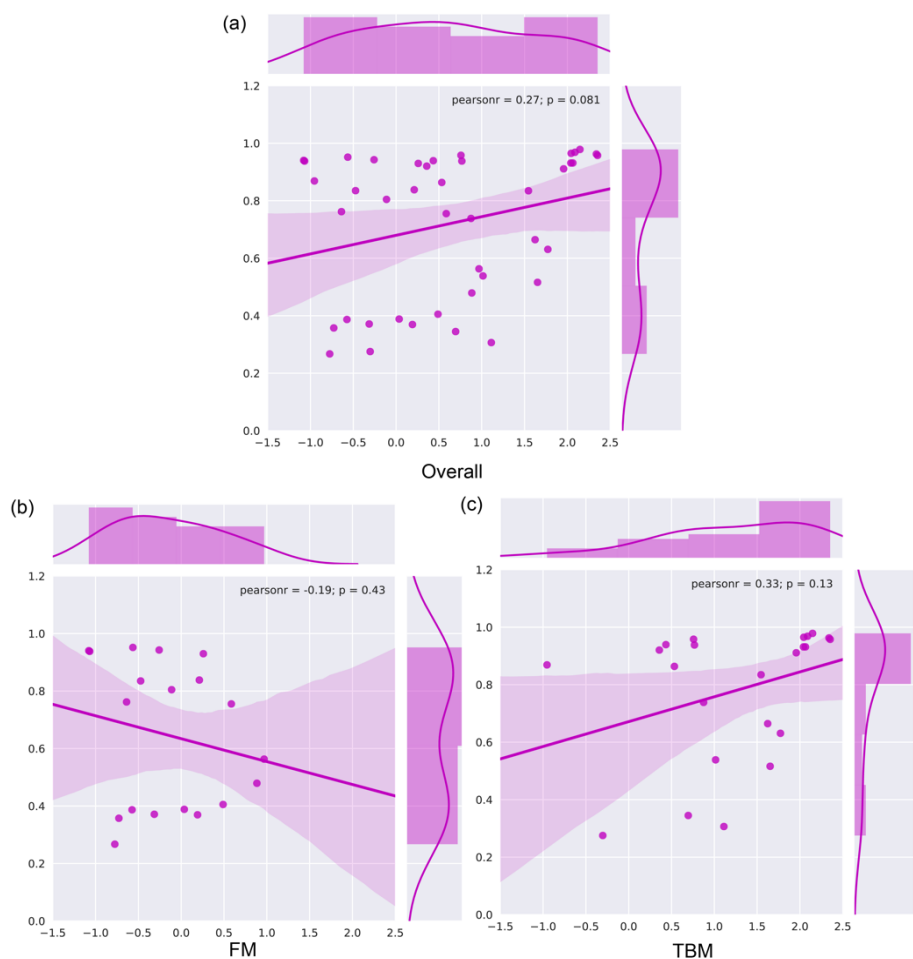

**Figure 6.** Correlation between model quality and alignment depth in MSA for our method tested in the CASP13 set. **(a)** All CASP13 targets. **(b)** CASP13 FM targets. **(c)** CASP13 TBM targets. The vertical axis represents the TM-score, while the horizontal axis represents the logarithm of alignment depth ($N/L$).

Structure prediction of membrane proteins is of very high value in practical usage since they are responsible for the material transport and signal transduction between cellular internal and external environments. Experimental structure determination is very hard for membrane proteins and therefore the data accumulation of known membrane protein structures is far from enough to support the regular training scheme of DNNs. However, the folding mechanism of all proteins should be the same for all proteins in the perspectives of physics and chemistry, which implies that good predictors may have good generalizability to allow the application on membrane proteins. We used 416 non-redundant membrane proteins from the PDBTM[23] set to test the generalizability of our method (**Tables S9** and **S12**). Without any transfer learning, our method can fold 57% of the proteins (236 out of 416) into the correct topology, achieving an average TM-score of 0.546. As an example, the chain A of target 5I20, an important exporter of Drug/Metabolite Transporter (DMT) superfamily in *E.coli.*, could be folded with very high accuracy (**Figure 7**). These results further confirm the applicability of our method on membrane proteins, although the models are trained mainly by protoplasmic soluble ones.

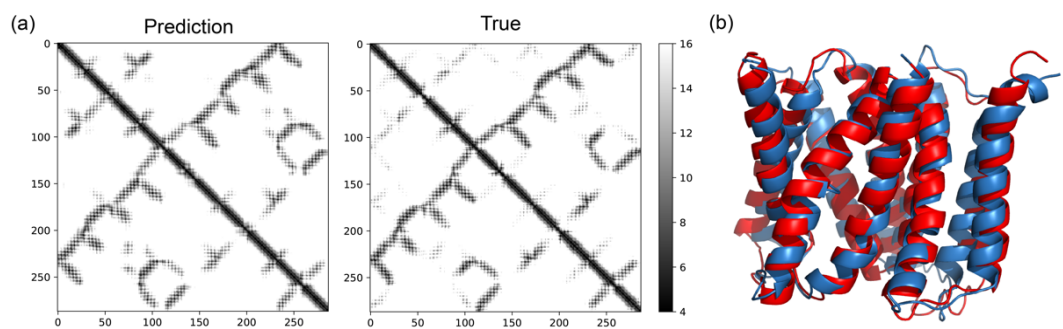

**Figure 7.** The prediction of our method on a membrane exporter (PDB ID: 5I20, chain A). **(a)** Comparison of the predicted distance map by our method (left) and true distance map (right). **(b)** Structure alignment of the best model folded using our method (red) against the crystal structure (blue).

## 3. Discussion

In this work, we treated protein residue distance prediction as a regression problem for the first time and precisely predicted continuous, real-valued distances merely from sequence information via an exquisitely designed GAN system. Through adversarial training procedure on the two parts of this system, i.e. the generator and the discriminator, our model could learn the texture pattern of protein residue distance map. Feeding these distance predictions to the basic CNS suite produces structures with competitive model quality compared with the state-of-the-art predictors. The structure quality predicted by our system has no significant correlation with the alignment depth in the MSA. Although trained by protoplasmic soluble proteins, the good generalizability ensures that it works for both soluble and membrane proteins.

The prediction power of our GAN system diminishes for long proteins (**Figure 5b** and **Table S15**), because the maximal protein length in our training procedure was set to 300 residues due to the memory limitation of our training facility. In addition, the generator of our GAN system is shallow, when compared with the 660-layer ResNet of AlphaFold. These limitations could be solved by training on more advanced hardware. Preliminary experiments on model 6 (see **Table S8**) via distributed training using 2 GPU cards show positive results: the absolute error (in the interval of 4-16 Å) for proteins of 300-400 residues and of > 400 residues (see the two sets in **Table S15**) are improved by 0.2 Å and 0.9 Å, respectively, on average.

We abandoned contact prediction because contact is a compromise when accurate distance prediction is not available. Real-valued distance has many advantages over contact, among which the most essential one is that a true real-valued distance map is a direct representation of a structure with all information included. Thus, developing differentiable distance-to-structure mapping functions that bridge our GAN system and the final structures will enable an end-to-end training procedure. Different from dihedral angle-based end-to-end differentiable system proposed earlier[24], which only considers local structure information of neighboring residues and fails at chirality, distance-based end-to-end differentiable system could extract global structure information for residue pairs with sequence separations of any length and determine the chirality because only one kind of chirality could be fitted with sufficient distance restrictions. In the future, our research interest would be such an end-to-end training scheme based on GAN system presented in this paper.

We have also noticed that a new kind of GAN, called cycle GAN, which provides a generalized semi-supervised learning approach for situations of large-scale label deficiency, has been proposed recently[25]. This method may further benefit distance prediction, since many proteins have known sequences but lack structures (label for prediction). With the rapid development of high-throughput sequencing technologies and the exponential data growth of protein sequences, especially the construction of metagenome databases, it is reasonable for us to believe that the new era of "protein structure determination via sequencing" would come in the near further.

## 4. Experimental Section

*Protein dataset*

All proteins in our training set were extracted from the SCOPe database of 2.05 version[26]. The cutoff of redundancy elimination was set as 20% sequence identity and the shortest protein of each protein family was picked out. The final training set contained 6862 proteins in total.

We evaluated our methods on three testing sets: the CASP12 set[3], the CASP13 set[4] and the PDBTM set[23] of membrane proteins (choosing only one chain for each protein target). Considering that proteins in our training set were all determined before the CASP12 and CASP13 competitions and that members in the PDBTM set are non-redundant to our training set, benchmarks on these testing sets could provide fair evaluation for our method.

*Feature generation*

The input features of our GAN system consist of 0D, 1D and 2D ones. First of all, the MSAs were built through HHblits[27] from the UniProt20 database[28]. The protein length and the alignment depth constitute the 0D features. The results of DeepCNF[29] and SPIDER3[30] together with the one-hot identities and appearing frequencies of amino acids at corresponding site in the MSAs constitute the 1D features. Co-evolution information extracted from the MSAs by CCMPred[31] and mutual information, together with relative position of every site to other sites and the amount of gaps in the MSAs for every site, constitute the 2D features. We broadcast 0D features and 1D features to match the shape of 2D features. Notably, the 1D features were broadcast twice, in the horizontal and vertical directions, which doubled the feature amounts. Finally, the 0D features (2 channels), 1D features (124 channels) and 2D features (4 channels) were concatenated as our input (130 channels in total).

*Mapping function*

To allow the real-valued distance regression via DNN with BN layers, we designed two different mapping functions for input features and labels, respectively. These mapping functions could map them into the interval of [-1, 1]. For simplicity, we call the space of the true values of features/labels as "real space" (RS) and the space of the mapped values as "training space" (TS), which is maintained only for training.

For every channel of the input features from RS to TS, we first calculated its maximum value and minimum value, and then mapped its elements uniformly via linear transformation:

$$V_{TS} = \frac{2V_{RS} - Max_{RS} - Min_{RS}}{Max_{RS} - Min_{RS}}, \tag{7}$$

where *V*, *Max* and *Min* denote current value, maximum and minimum of this channel, respectively, and the subscripts represent the corresponding spaces.

Our attention for label (i.e. ground-truth distance) transformation was mainly on the interval of 4-16 Å, since distances within this interval are the most valuable ones for protein folding. To disperse values within this interval to the utmost and to take all values into consideration instead of setting a cutoff and throwing some away, we chose *tanh* as our mapping function here. Before using *tanh*, we did a linear transformation that could map from the interval of [4, 16] to that of [-2.5, 2.5], where *tanh* has large first derivatives. The label mapping function is

$$V_{TS} = tanh\left(\frac{5 \times V_{RS} - 50}{12}\right). \tag{8}$$

Since it is an invertible function, we could use its inverse to derive the final distance prediction from the outputs of DNN.

*Attention module*

We implemented this module in two steps, with the first focusing on the attention of channels, and after its multiplication with the raw inputs, the second one focusing on the spatial attention of pixels. The final inputs after the process of the attention module would be:

$$Inputs = (Raw \times CAF(Raw)) \times PAF(Raw \times CAF(Raw)), \qquad (9)$$

where *CAF* and *PAF* represent channel-wised attention function and pixel-wised attention function, respectively. In the first step (CAF), we used a 3-layer perceptron with a bottleneck architecture (i.e. 130-75-130) to process the summation of the global average and max pooling results for individual input channels. This architecture is not only light-weighted but also effective for information integration. Imaging to mix a bottle of half juice and half water, the most effective way is to squeeze the bottle neck and then loose it. ReLU activation was used in all layers except the last one, which used Sigmoid activation to force channel weights to fall in the interval of [0, 1]. In the second step (PAF), we used one single 7×7 convolution filter with stride 1 to scan the concatenation of the global average and max pooling results of individual input pixels and also the Sigmoid function to output pixel weights. The convolution kernel would determine whether the current pixel is an important one or not according to its neighboring zone.

*MD simulations*

For all proteins in our training set, MD simulations were conducted in a water box with periodic boundary conditions (PBC) applied. The edge length of the water box was 10 Å larger than the diameter of the protein, and its volume was about 354141.7 Å$^3$ on average. To simulate the physiological environment, 160 mmol/L NaCl was added into the system (~ 37 Na$^+$ and Cl$^-$ ions in the water box). The specific amounts of Na$^+$ and Cl$^-$ ions were set slightly different for each individual protein to ensure the electric neutrality of the system. The simulation procedure contained 3 stages: energy minimization, system heating and equilibrium simulation. In the energy minimization stage, the Low-MODe (LMOD) method was employed for 5000 steps, in which the steepest descent was applied for the first 2500 steps and then switched to the conjugate gradient descent from the next 2500 steps. In the heating stage, the volume of the system was fixed and a total of 8000 time steps were conducted with the step size of 2 femtoseconds, during which the temperature of the system was gradually heated up to 300 K from 0 K in the first 6000 time steps and was maintained at 300 K for the following 2000 time steps. In the final equilibrium simulation stage, canonical ensemble (NVT) was adopted and the simulation was run for 2,500,000 time steps, i.e. 5 nanoseconds, in total. Bond interactions involving H-atoms were fixed in the last 2 stages. We saved one structure every 5000 time steps from the simulation trajectory.

**Authors' contributions**

W. D. and H. G. proposed the initial idea and designed the methodology. W. D.

implemented the concept and processed the results. W. D. and H. G. wrote the manuscript. Both authors read and approved the final manuscript.

**Supplementary figures**

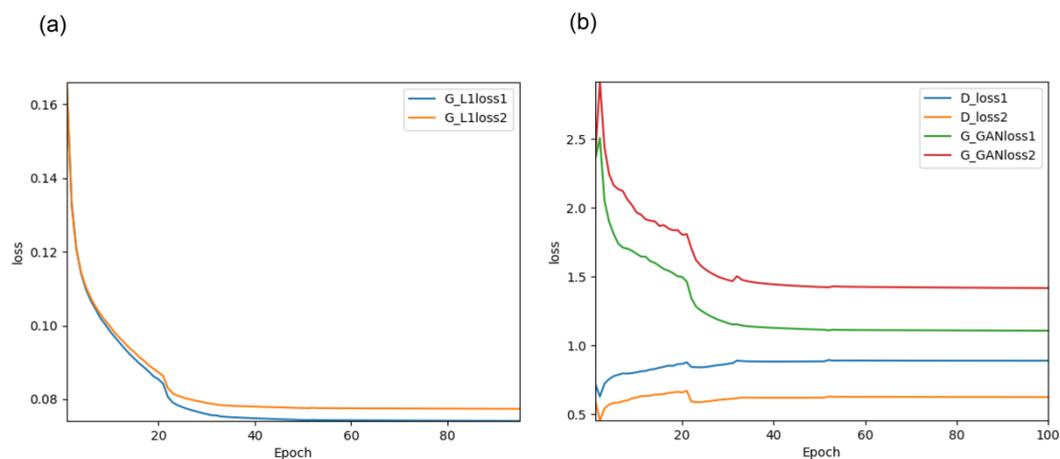

**Figure S1.** Patch classifiers in D help to improve the model training. The patch classifier of D (implemented as FCN) is marked by the suffix of "1"and the original single classifier (implemented as SPP) that makes judgement on the entire input is marked by the suffix of "2". **(a)** L1 loss of G. **(b)** Loss of D and GAN loss of G. Generally, for a GAN system, the smaller GAN loss of G and larger loss of D both mean that G is more likely to produce reality-indistinguishable fakes to fool D. Thus, the simultaneous reduction of GAN loss of G and enhancement of D loss in the case "1" imply that the distance map produced by G becomes more realistic when the patch classifier is applied.

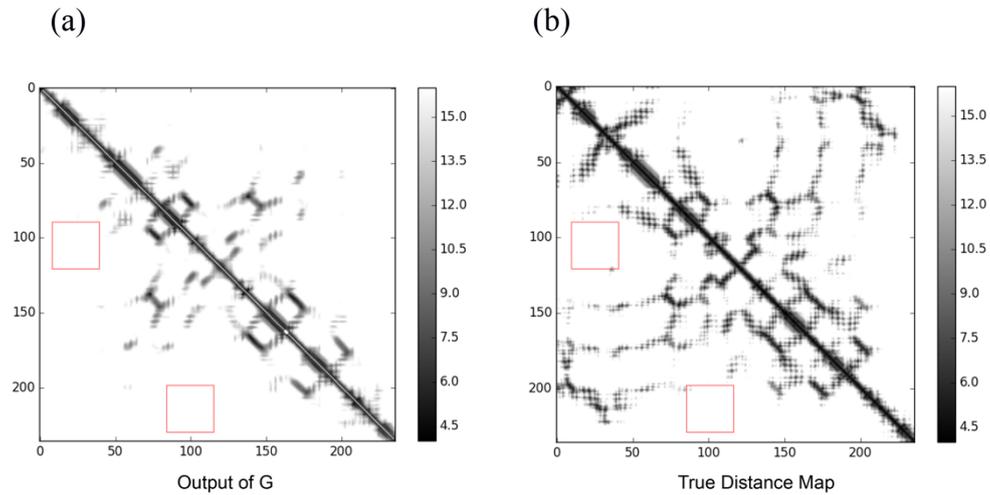

**Figure S2.** Patches of the blank regions that lack contributive information for folding may cause confusion in the judgement of D. The premature prediction produced by G in the early stage of GAN training (**a**) and the true distance map (**b**) are shown as an example here. 34×34 patches marked by red squares are likely to be identified as true even if many stripes with intense signals are completely missed in (**a**).

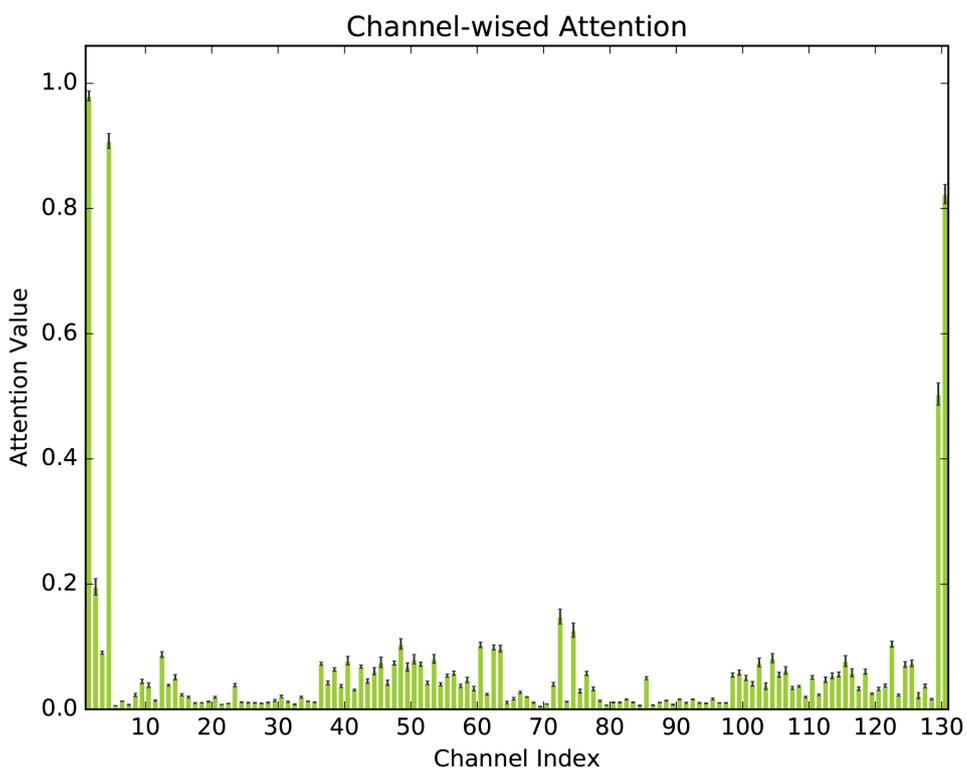

**Figure S3.** Channel-wised attention. Error bars are added on the head of each pillar. Our input feature has 130 channels in total (see Experimental Section), in which the first four (indices 1-4, corresponding to CCMPred, MI, MSA gap frequency and relative residue position) are 2D features, the latter two (indices 129 and 130, corresponding to target length and MSA count) are 0D features and the rest are broadcast 1D features. Signal strength of 1D features is inhibited dramatically by the channel-wised attention operation.

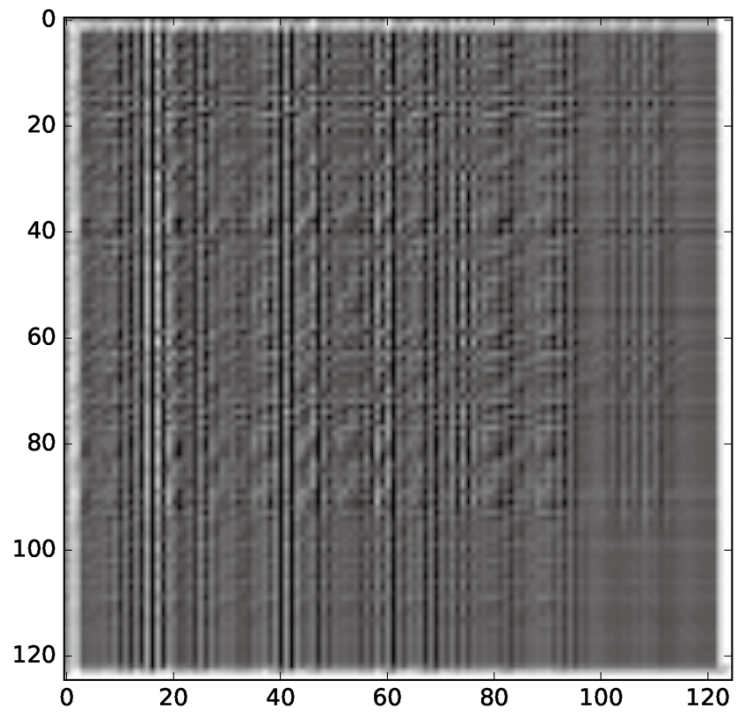

**Figure S4.** An example of the pixel-wised attention map. Clearly, the weight of each pixel (shown in the grayscale) has been readjusted after the pixel-wised attention operation.

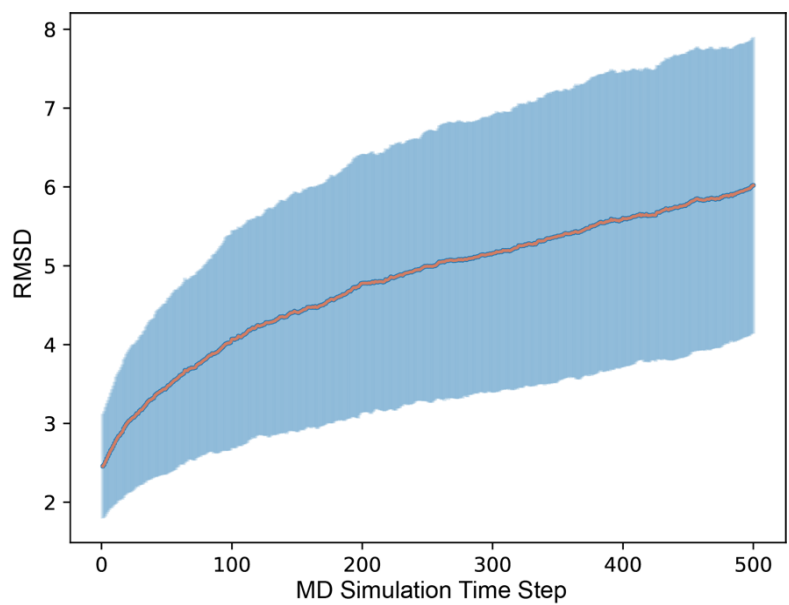

**Figure S5.** Structure oscillation in MD simulations. The orange line represents the average RMSD of all proteins in our training set along the simulation time step. The blue shadow represents corresponding error bar at each time step.

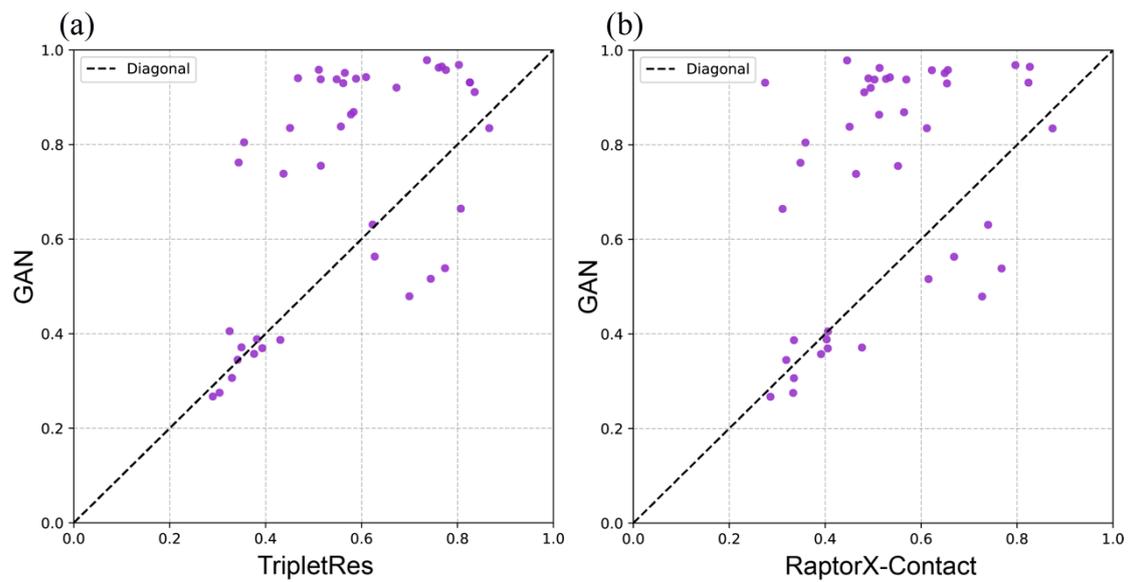

**Figure S6.** Comparison of TM-score for structures modelled using predictions of our GAN system as constraints against those using results of the top contact prediction groups in CASP 13. **(a)** Our GAN vs. TripletRes. **(b)** Our GAN vs. RaptorX-Contact. The vertical and horizontal axes denote the TM-scores produced by our method and by the compared contact predictors, respectively.

**Supplementary tables**

**Table S1.** Performance of the ResNet and cGAN models in the validation set.

| Predicted distance interval (Å) | ResNet | cGAN |
|:---:|:---:|:---:|
| 4-6 | 0.641 | 0.999 |
| 6-8 | 0.871 | 1.051 |
| 8-10 | 1.464 | 1.981 |
| 10-12 | 1.721 | 2.202 |
| 12-14 | 2.225 | 2.075 |
| 14-16 | 2.557 | 2.278 |
| Overall (4-16) | 1.832 | 1.938 |

Here, we used the mean absolute error (Å) to evaluate the prediction errors for residue pairs with predicted distances falling within various intervals.

**Table S2.** Model performance for different architectures of G.

| | Interval of predicted distance (Å) | | | | | | |
|---|---|---|---|---|---|---|---|
| | 4-6 | 6-8 | 8-10 | 10-12 | 12-14 | 14-16 | 4-16 |
| ResNet: 2 layers per block | 0.973 | 1.087 | 1.952 | 2.136 | 2.109 | 2.281 | 1.920 |
| **ResNet: 3 layers per block (bottleneck channel)** | **0.755** | **0.915** | **1.412** | **1.735** | **1.889** | **2.105** | **1.640** |
| ResNet: 3 layers per block (barrel-like channel) | 0.729 | 0.952 | 1.414 | 1.848 | 2.061 | 2.287 | 1.754 |
| ResNet: Bottleneck | 0.891 | 1.048 | 1.726 | 2.218 | 2.324 | 2.550 | 2.011 |
| U-Net | 1.660 | 1.510 | 1.884 | 4.047 | 4.865 | 5.180 | 4.334 |
| Dense Net | 0.859 | 1.265 | 2.184 | 2.879 | 3.155 | 3.455 | 2.696 |

Model performance is quantified by the mean absolute error (Å) in the 5-fold cross validation. For fair comparison, all architectures were designed with approximately the same computational consumption (~ 5800MiB FLOPs). The picked architecture is marked in bold.

**Table S3.** Model performance for different kernel sizes of G.

|  | Interval of predicted distance (Å) | | | | | | |
| --- | --- | --- | --- | --- | --- | --- | --- |
|  | 4-6 | 6-8 | 8-10 | 10-12 | 12-14 | 14-16 | 4-16 |
| 3×3 | 0.736 | 0.859 | 1.253 | 1.547 | 1.672 | 1.894 | 1.473 |
| 5×5 | 0.726 | 0.822 | 1.109 | 1.381 | 1.524 | 1.760 | 1.351 |
| **7×7** | **0.717** | **0.816** | **1.034** | **1.274** | **1.426** | **1.667** | **1.271** |
| 3×3 with dilation rate of 2 | 1.011 | 0.969 | 1.425 | 1.644 | 1.762 | 1.999 | 1.571 |
| Separable convolution kernel with size of 3 | 0.805 | 0.912 | 1.370 | 1.729 | 1.844 | 2.040 | 1.607 |

Model performance is quantified by the mean absolute error (Å) in the 5-fold cross validation. The networks compared here are completely the same except the kernels. The picked kernel is marked in bold.

**Table S4.** Model performance for different activation functions of G.

|  | Interval of predicted distance (Å) | | | | | | |
| --- | --- | --- | --- | --- | --- | --- | --- |
|  | 4-6 | 6-8 | 8-10 | 10-12 | 12-14 | 14-16 | 4-16 Å |
| Leaky-ReLU | 0.736 | 0.859 | 1.253 | 1.547 | 1.672 | 1.894 | 1.473 |
| ELU | 0.715 | 0.825 | 1.181 | 1.445 | 1.570 | 1.817 | 1.392 |
| C-ReLU | 0.815 | 0.974 | 1.375 | 1.757 | 1.898 | 2.115 | 1.656 |
| P-ReLU | 0.702 | 0.808 | 1.180 | 1.440 | 1.574 | 1.824 | 1.395 |
| R-ReLU | 1.635 | 1.614 | 1.676 | 2.441 | 2.864 | 3.164 | 2.561 |
| tanh | 0.862 | 1.265 | 2.183 | 2.876 | 3.153 | 3.450 | 2.693 |
| Softplus | 0.690 | 0.792 | 1.152 | 1.391 | 1.530 | 1.789 | 1.361 |
| Softsign | 1.045 | 1.235 | 1.940 | 2.529 | 2.595 | 2.823 | 2.264 |
| Swish | 0.688 | 0.792 | 1.147 | 1.401 | 1.521 | 1.780 | 1.357 |
| **Swish (with parameter)** | **0.670** | **0.790** | **1.131** | **1.388** | **1.509** | **1.751** | **1.340** |

Model performance is quantified by the mean absolute error (Å) in the 5-fold cross validation. The networks compared here are completely the same except the activations. The picked activation is marked in bold.

**Table S5.** Model performance for different loss functions of G.

|  | Interval of predicted distance (Å) | | | | | | |
| --- | --- | --- | --- | --- | --- | --- | --- |
|  | 4-6 | 6-8 | 8-10 | 10-12 | 12-14 | 14-16 | 4-16 |
| **L1 Loss** | **0.736** | **0.859** | **1.253** | **1.547** | **1.672** | **1.894** | **1.473** |
| L2 Loss | 1.482 | 1.329 | 2.370 | 2.548 | 3.075 | 5.372 | 3.766 |
| Huber Loss | 1.310 | 1.307 | 2.083 | 2.316 | 2.318 | 3.115 | 2.330 |
| Logcosh loss | 1.132 | 1.060 | 1.440 | 1.751 | 2.035 | 2.961 | 1.986 |
| LogMSE loss | 1.878 | 1.657 | 2.104 | 1.944 | 2.467 | 3.540 | 2.482 |
| PercentageMAE loss | - | - | - | - | - | - | - |

Model performance is quantified by the mean absolute error (Å) in the 5-fold cross validation. The networks compared here are completely the same except the losses. It is noteworthy that for Huber loss, we did a series of experiments at different values of its parameter δ and listed the best performing one here. "-" means training failure. The picked loss function is marked in bold.

**Table S6.** Model performance for different attention adding strategies of G.

| | Interval of predicted distance (Å) | | | | | | |
|---|---|---|---|---|---|---|---|
| | 4-6 | 6-8 | 8-10 | 10-12 | 12-14 | 14-16 | 4-16 |
| Without attention module | 0.736 | 0.859 | 1.253 | 1.547 | 1.672 | 1.894 | 1.473 |
| **Attention module on input only** | **0.704** | **0.787** | **1.139** | **1.425** | **1.547** | **1.749** | **1.351** |
| Attention module in each block | 0.827 | 0.915 | 1.388 | 1.622 | 1.698 | 1.841 | 1.501 |
| Attention module in each layer | 1.988 | 1.859 | 2.873 | 4.339 | 4.622 | 4.572 | 3.943 |

Model performance is quantified by the mean absolute error (Å) in the 5-fold cross validation. The networks compared here are completely the same except the attention adding strategies. The picked strategy is marked in bold.

**Table S7.** Comparison of the average TM-score for structures modelled using predictions of two same-architecture models trained by different-sourced labels.

|  | Model 1 | Model 2 |
|---|---|---|
| CASP13 set (42 targets) | 0.579 | 0.619 |
| Membrane protein set (416 targets) | 0.500 | 0.516 |

Model 1 is trained using crystal structures directly from PDB. Model 2 is trained by structures produced by MD simulations. For both models, the architecture is identical to model 3 in Table S8.

**Table S8.** The 14 models selected for boosting the variation of prediction results.

| Model index | Weight of L1 loss in the loss of G | Patch size of D | Clipping in the D loss |
|---|---|---|---|
| 1 | 158 | 34 | No |
| 2 | 258 | 34 | No |
| 3 | 100 | 70 | No |
| 4 | 158 | 70 | No |
| 5 | 200 | 70 | No |
| 6 | 258 | 70 | No |
| 7 | 158 | 142 | No |
| 8 | 258 | 142 | No |
| 9 | 258 | 34 | Yes |
| 10 | 100 | 70 | Yes |
| 11 | 158 | 70 | Yes |
| 12 | 200 | 70 | Yes |
| 13 | 258 | 70 | Yes |
| 14 | 258 | 142 | Yes |

**Table S9.** Overall performance on CASP12, CASP13 and membrane-protein datasets.

| Model index | CASP12 | | | CASP13 | | | Membrane Proteins | | |
|---|---|---|---|---|---|---|---|---|---|
| | Precision | Recall | F1 | Precision | Recall | F1 | Precision | Recall | F1 |
| 1 | 0.736 | 0.725 | 0.727 | 0.787 | 0.793 | 0.786 | 0.699 | 0.774 | 0.726 |
| 2 | 0.764 | 0.699 | 0.727 | 0.755 | 0.718 | 0.731 | 0.723 | 0.751 | 0.728 |
| 3 | 0.757 | 0.704 | 0.727 | 0.797 | 0.769 | 0.779 | 0.716 | 0.752 | 0.725 |
| 4 | 0.743 | 0.718 | 0.727 | 0.736 | 0.736 | 0.731 | 0.714 | 0.775 | 0.735 |
| 5 | 0.763 | 0.694 | 0.724 | 0.789 | 0.758 | 0.772 | 0.757 | 0.726 | 0.732 |
| 6 | 0.759 | 0.698 | 0.724 | 0.743 | 0.823 | 0.778 | 0.723 | 0.739 | 0.722 |
| 7 | 0.739 | 0.696 | 0.712 | 0.733 | 0.715 | 0.719 | 0.690 | 0.751 | 0.710 |
| 8 | 0.775 | 0.699 | 0.731 | 0.759 | 0.719 | 0.733 | 0.726 | 0.747 | 0.728 |
| 9 | 0.803 | 0.675 | 0.729 | 0.787 | 0.710 | 0.743 | 0.795 | 0.730 | 0.751 |
| 10 | 0.769 | 0.696 | 0.727 | 0.759 | 0.712 | 0.730 | 0.728 | 0.749 | 0.729 |
| 11 | 0.775 | 0.703 | 0.733 | 0.758 | 0.715 | 0.730 | 0.735 | 0.773 | 0.745 |
| 12 | 0.818 | 0.658 | 0.724 | 0.834 | 0.695 | 0.753 | 0.796 | 0.710 | 0.740 |
| 13 | 0.791 | 0.693 | 0.735 | 0.788 | 0.707 | 0.742 | 0.770 | 0.745 | 0.748 |
| 14 | 0.786 | 0.693 | 0.734 | 0.777 | 0.712 | 0.739 | 0.765 | 0.752 | 0.750 |

Precision here is the percentage of residue pairs with predicted distances in the 4-16 Å interval that indeed have true distances in that interval. Recall here is the percentage of residue pairs with true distances in the 4-16 Å interval that are also predicted in that interval. F1 here is calculated as the harmonic mean of precision and recall, i.e.

$F1 = 2(precision \times recall)/(precision + recall)$.

**Table S10.** Model performance on the CASP12 dataset.

| Model index | Absolute error (Å) | | | | | | | Relative error (%) 4-16 Å |
|---|---|---|---|---|---|---|---|---|
| | 4-6 Å | 6-8 Å | 8-10 Å | 10-12 Å | 12-14 Å | 14-16 Å | 4-16 Å | |
| 1 | 1.080 | 1.633 | 2.818 | 2.815 | 3.471 | 3.789 | 3.002 | 22.5 |
| 2 | 0.954 | 1.498 | 2.533 | 2.683 | 3.287 | 3.583 | 2.778 | 22.3 |
| 3 | 1.402 | 1.542 | 2.525 | 2.617 | 3.224 | 3.622 | 2.813 | 22.3 |
| 4 | 1.029 | 1.728 | 2.734 | 2.771 | 3.430 | 3.728 | 2.952 | 22.4 |
| 5 | 0.800 | 1.176 | 2.444 | 3.180 | 3.308 | 3.658 | 2.867 | 23.0 |
| 6 | 0.852 | 1.469 | 2.502 | 2.707 | 3.419 | 3.751 | 2.885 | 22.8 |
| 7 | 0.945 | 1.481 | 2.643 | 2.860 | 3.601 | 3.979 | 3.026 | 23.4 |
| 8 | 0.730 | 1.341 | 2.360 | 2.529 | 3.150 | 3.474 | 2.612 | 21.8 |
| 9 | 0.644 | 1.156 | 1.973 | 2.086 | 2.786 | 3.236 | 2.313 | 20.9 |
| 10 | 0.743 | 1.363 | 2.337 | 2.541 | 3.262 | 3.640 | 2.702 | 22.2 |
| 11 | 0.823 | 1.380 | 2.373 | 2.492 | 3.081 | 3.394 | 2.596 | 21.6 |
| 12 | 0.539 | 1.054 | 1.853 | 2.020 | 2.688 | 3.138 | 2.218 | 21.2 |
| 13 | 0.660 | 1.264 | 2.172 | 2.276 | 2.899 | 3.226 | 2.405 | 21.0 |
| 14 | 0.710 | 1.292 | 2.235 | 2.365 | 3.040 | 3.327 | 2.502 | 21.3 |

Absolute error here is the absolute difference (Å) between predicted distance and its corresponding true distance. Relative error here is the absolute error normalized by its corresponding true distance.

**Table S11.** Model performance on the CASP13 dataset.

| Model index | Absolute error (Å) | | | | | | | Relative error (%) 4-16 Å |
|---|---|---|---|---|---|---|---|---|
| | 4-6 Å | 6-8 Å | 8-10 Å | 10-12 Å | 12-14 Å | 14-16 Å | 4-16 Å | |
| 1 | 0.989 | 1.478 | 2.200 | 2.201 | 2.735 | 3.205 | 2.509 | 18.0 |
| 2 | 1.068 | 1.665 | 2.689 | 2.889 | 3.492 | 3.913 | 3.053 | 23.3 |
| 3 | 1.357 | 1.384 | 2.071 | 2.119 | 2.578 | 3.136 | 2.412 | 17.8 |
| 4 | 1.152 | 1.883 | 2.874 | 2.938 | 3.610 | 4.058 | 3.234 | 23.3 |
| 5 | 0.886 | 1.176 | 2.400 | 2.899 | 2.921 | 3.173 | 2.611 | 19.7 |
| 6 | 0.855 | 1.463 | 2.443 | 2.675 | 3.382 | 3.691 | 2.928 | 19.2 |
| 7 | 1.063 | 1.473 | 2.747 | 3.064 | 3.714 | 4.099 | 3.212 | 23.8 |
| 8 | 0.822 | 1.460 | 2.637 | 2.865 | 3.517 | 3.865 | 2.997 | 23.0 |
| 9 | 0.770 | 1.400 | 2.367 | 2.517 | 3.114 | 3.569 | 2.685 | 22.0 |
| 10 | 0.816 | 1.500 | 2.532 | 2.798 | 3.520 | 3.948 | 3.021 | 23.4 |
| 11 | 0.929 | 1.496 | 2.579 | 2.729 | 3.336 | 3.725 | 2.903 | 22.8 |
| 12 | 0.664 | 1.294 | 2.213 | 2.376 | 3.043 | 3.567 | 2.640 | 19.6 |
| 13 | 0.626 | 1.254 | 2.213 | 2.338 | 2.967 | 3.466 | 2.554 | 21.9 |
| 14 | 0.762 | 1.404 | 2.397 | 2.529 | 3.221 | 3.584 | 2.747 | 22.3 |

Absolute error here is the absolute difference (Å) between predicted distance and its corresponding true distance. Relative error here is the absolute error normalized by its corresponding true distance.

**Table S12.** Model performance on the non-redundant membrane protein dataset.

| Model index | Absolute error (Å) | | | | | | | Relative error (%) 4-16 Å |
|---|---|---|---|---|---|---|---|---|
| | 4-6 Å | 6-8 Å | 8-10 Å | 10-12 Å | 12-14 Å | 14-16 Å | 4-16 Å | |
| 1 | 1.217 | 1.658 | 2.813 | 2.862 | 3.773 | 4.431 | 3.349 | 21.6 |
| 2 | 1.060 | 1.593 | 2.732 | 2.940 | 3.858 | 4.354 | 3.232 | 21.6 |
| 3 | 1.507 | 1.609 | 2.676 | 2.790 | 3.559 | 4.222 | 3.166 | 21.5 |
| 4 | 1.153 | 1.665 | 2.702 | 2.784 | 3.643 | 4.305 | 3.241 | 21.0 |
| 5 | 0.840 | 1.135 | 2.697 | 3.335 | 3.364 | 3.775 | 2.955 | 21.0 |
| 6 | 0.971 | 1.500 | 2.647 | 2.862 | 3.926 | 4.531 | 3.315 | 21.8 |
| 7 | 1.036 | 1.397 | 2.865 | 3.327 | 4.240 | 4.803 | 3.575 | 23.1 |
| 8 | 0.818 | 1.357 | 2.632 | 2.968 | 3.954 | 4.333 | 3.248 | 21.7 |
| 9 | 0.579 | 1.051 | 1.864 | 2.016 | 2.753 | 3.301 | 2.302 | 18.4 |
| 10 | 0.744 | 1.314 | 2.408 | 2.794 | 3.865 | 4.473 | 3.178 | 21.3 |
| 11 | 0.947 | 1.403 | 2.420 | 2.635 | 3.435 | 3.782 | 2.855 | 19.8 |
| 12 | 0.578 | 0.993 | 1.797 | 1.985 | 2.781 | 3.456 | 2.349 | 19.3 |
| 13 | 0.643 | 1.184 | 2.105 | 2.291 | 3.062 | 3.523 | 2.512 | 19.2 |
| 14 | 0.700 | 1.199 | 2.141 | 2.399 | 3.214 | 3.636 | 2.634 | 19.3 |

Absolute error here is the absolute difference (Å) between predicted distance and its corresponding true distance. Relative error here is the absolute error normalized by its corresponding true distance.

**Table S13.** Comparison of structures modelled using predictions of our GAN system as constraints against those using DeepConPred2 on CASP12 targets.

|  | Average TM Score | Correctly Folded |
|---|---|---|
| DeepConPred2 | 0.409 | 10 |
| Our GAN system | 0.661 | 26 |

"Correctly Folded" means the number of targets folded in the correct topology (TM-score > 0.5).

**Table S14.** Comparison of structures modelled using predictions of our GAN system as constraints against the top groups in CASP 13.

|  | Overall (39 targets) | | FM (19 targets) | | TBM (20 targets) | |
| --- | --- | --- | --- | --- | --- | --- |
|  | TM-score | Correctly Folded | TM-score | Correctly Folded | TM-score | Correctly Folded |
| A7D | 0.706 | 33 | 0.633 | 14 | 0.776 | 19 |
| Zhang | 0.689 | 33 | 0.554 | 14 | 0.818 | 19 |
| MULTICOM | 0.681 | 32 | 0.557 | 13 | 0.799 | 19 |
| Our method | 0.701 | 28 | 0.646 | 11 | 0.753 | 17 |

TM-score is averaged over the targets tested. "Correctly Folded" means the number of targets folded in the correct topology (TM-score > 0.5).

**Table S15.** Model performance for proteins of different sizes.

| | Absolute error (Å) | | | | | | | Relative error (%) 4-16 Å |
|---|---|---|---|---|---|---|---|---|
| | 4-6 Å | 6-8 Å | 8-10 Å | 10-12 Å | 12-14 Å | 14-16 Å | 4-16 Å | |
| Validation Set 1 (300-400 residues) | 0.809 | 1.188 | 1.914 | 2.065 | 2.670 | 3.022 | 2.241 | 0.178 |
| Validation Set 2 (>400 residues) | 1.818 | 2.148 | 3.903 | 4.406 | 5.442 | 6.016 | 4.836 | 0.273 |

We picked out proteins with length between 300 and 400 (450 proteins) to construct the validation set 1 and proteins longer than 400 (249 proteins) to construct the validation set 2. Absolute error here is the absolute difference (Å) between predicted distance and its corresponding true distance. Relative error here is the absolute error normalized by its corresponding true distance.